\documentclass[%
  english,
  reprint,
  superscriptaddress,
  amsmath,amssymb,
  aps,
  pra,
  floatfix,
]{revtex4-2}

\usepackage{graphicx}
\usepackage[tight]{subfigure}
\usepackage{dcolumn}
\usepackage{bm}
\usepackage[colorlinks,linkcolor={blue},citecolor={blue},urlcolor={blue}]{hyperref}
\usepackage[table]{xcolor}
\usepackage{multirow}
\usepackage{array}
\usepackage{booktabs}
\usepackage{ctable}
\usepackage{upgreek}
\usepackage{epsfig}
\usepackage{mathrsfs}
\usepackage{amssymb}
\usepackage{amsbsy}
\usepackage{color}
\usepackage{cancel}
\usepackage{pifont}
\usepackage{marginnote}
\usepackage{dsfont}
\usepackage{babel}

\newcommand{\abs}[1]{\lvert#1\rvert}

\newcommand{\bcen}{\begin{center}}
\newcommand{\ecen}{\end{center}}
\newcommand{\btab}{\begin{tabular}}
\newcommand{\etab}{\end{tabular}}
\newcommand{\bdes}{\begin{description}}
\newcommand{\edes}{\end{description}}

\newcommand{\beq}{\begin{equation}}
\newcommand{\eeq}{\end{equation}}
\newcommand{\bea}{\begin{eqnarray}}
\newcommand{\eea}{\end{eqnarray}}

\newcommand{\bary}{\begin{array}}
\newcommand{\eary}{\end{array}}
\newcommand{\benum}{\begin{enumerate}}
\newcommand{\eenum}{\end{enumerate}}
\newcommand{\bitem}{\begin{itemize}}
\newcommand{\eitem}{\end{itemize}}

\newcommand{\mean}[1]{\langle #1 \rangle}

\newcommand{\Eqn}[1] {Eq.~(\ref{#1})}

\newcommand{\Sect}[1] {Section~\ref{#1}}

\newcommand{\Fig}[1]{Fig.~\ref{#1}}

\newcommand{\Tabs}[1]{Tab.~\ref{#1}}

\newcommand{\opt}[1]{}

\newcommand{\titlename}{Interplay of disorder and interactions
  in the bilayer band-insulator : A determinant quantum Monte Carlo study}

\begin{document}


\title{\titlename}

\author{Yogeshwar Prasad}
\email{yogeshwar2609@kangwon.ac.kr}
\affiliation{Center for Condensed Matter Theory,
  Department of Physics, Indian Institute of Science,
  Bangalore 560012, India}
\affiliation{Department of Liberal Studies,
  Kangwon National University,
  Samcheok, 25913, Republic of Korea}

\author{Hunpyo Lee}
\email{hplee@kangwon.ac.kr}
\affiliation{Department of Liberal 
Studies,
  Kangwon National University,
  Samcheok, 25913, Republic of Korea}

\date{\today}

\begin{abstract}

In previous studies of the half-filled bilayer attractive Hubbard model
[Prasad {\it et al.} Phys. Rev. A {\bf 89}, 043605 (2014);
Prasad, Phys. Rev. B {\bf 106}, 184506 (2022)], it has
been shown that the clean system has a band-insulator (BI) to superfluid (SF)
quantum phase transition. In this paper, we append the effects of random on-site
disorder on the kinetic energy, double occupancy, and the pair-pair correlations
in the bilayer model. Using the determinant quantum Monte Carlo simulation, we
observe that the on-site random disorder plays a significant role in the
localization of on-site pairs, and hence in the reduction of the effective hopping.
This results in an increase of the double occupancy, which is an effect that is similar
to the attractive interaction. We find no change in the critical value of the
interaction at which the model undergoes a transition from the BI to SF regime,
even though the pair-pair correlations get suppressed for finite on-site disorder
strengths $V_d/t=0.1-0.8$. We also confirm that the weak-disorder suppresses the SF phase
largely in the strong-coupling limit. Hence the region of the SF phase reduces
in the presence of random on-site disorder. Finally, through finite-size scaling,
we have estimated the critical disorder strength $V_d^c/t\sim 1.44$ at
$\abs{U}/t=5$.

\end{abstract}


\maketitle


\section{Introduction}

  Anderson \cite{Anderson1959} argued that the disorder in the
absence of any interaction leads to the localization of the
electronic states. On the other hand, the attractive interaction
between the electrons lead to superconductivity, a very good
example of the long-range order in physics. The competition
between the superconductivity and the localization raises
profound questions in condensed matter physics. The interplay of
the effects of the interactions and the localization results in
the destruction of the superconductivity with an increasing
disorder and this leads to the superconductor–insulator (SI)
\cite{Trivedi1996,Scalettar1999} or superconductor–metal 
transition \cite{Dynes1984,White1986}. Earlier it was recognized
that $s$-wave superconductivity is remarkably robust against weak
disorder \cite{Anderson1959,Abrikosov1959}. It has also been argued
that the superfluid (SF) phase can survive even when single-particle states are
localized by disorder \cite{Ma1985}. Also, the $s$-wave superconductivity
in two coupled Hubbard chains is more resistant to disorder than in
the one-chain case \cite{Orignac1996}.
In spite of the decades of research, a generally accepted physical
picture of how the SF state is destroyed and the nature of the SI
transition have not yet been understood. Ultracold atomic gases in
optical lattices offer an opportunity to emulate these fundamental
issues. The disorder in an optical lattice can be introduced
\cite{Paiva2015} using the optical speckle
\cite{Damski2003,Lye2005,Clement2006,Pasienski2010,Billy2008},
impurities \cite{Gadway2011}, or a quasiperiodic optical lattice
\cite{Fallani2007,Roati2008}.

  Motivated by the recent developments in the realization of an
attractive Hubbard model on optical lattices \cite{Mitra2018,Gall2020},
we investigate the interplay between the on-site random disorder and
the attractive interaction on the long-range pair-pair correlations
in the two-dimensional bilayer band-insulator (BI) model at half filling.
The bilayer BI model has been studied earlier in the absence of any
disorder, both with on-site attractive Hubbard interaction
\cite{Prasad2014,Prasad2022} and repulsive Hubbard interactions
\cite{Lee2014,Ruger2014}, via quantum Monte Carlo and cluster dynamical
mean-field approaches. The quantum Monte Carlo studies for the disordered
attractive Hubbard model have been done in the past for a single-layer square
lattice \cite{Huscroft1997,Scalettar1999}. It has been found that in the
single-layer attractive Hubbard model (AHM) at half filling in a square lattice,
the superconducting order survives randomness out to a critical amount of
disorder, but the charge ordering state is immediately destroyed \cite{Huscroft1997}.

  In this work, we employ the exact and unbiased determinant quantum Monte
Carlo (DQMC) technique to study the two-particle properties such as pair
correlations in the bilayer BI (discussed in Refs. \cite{Prasad2014,Prasad2022})
with random on-site disorder. The paper is organized as follows: In \Sect{sec:model},
we briefly describe the bilayer BI model in the presence of the attractive Hubbard
interaction with random on-site disorder. We also discuss the computational details
of the DQMC technique used to investigate the model. In \Sect{sec:Results}, we
investigate the effect of disorder on the double occupancy and the effective
hopping, and on the two-particle pair-pair correlations. We also compare our
results with the clean system. We find that the pair-pair correlation
survives in the weak-disorder limit. The weak disorder suppresses the SF phase
largely in the strong-coupling limit, whereas the effect of disorder on the
pair-pair correlations is minimal in the weak-coupling limit. We perform scaling
analysis to estimate the critical disorder strength required to destroy the SF
phase. Finally, we conclude by providing a phase diagram after summarizing our
results in \Sect{sec:Concl}.



\section{Model and Computational Method}
\label{sec:model}

\subsection{Disordered bilayer band-insulator model}

  We start with a band-insulating state in the absence of any interaction
such that the hoppings in both the layers of the bilayer square lattice
are with opposite signs and the band gap is determined by the intralayer
hopping, as studied in Refs. \cite{Prasad2014,Prasad2022}. The Hamiltonian
of the system in the presence of on-site random disorder is
\begin{eqnarray}
  \label{eqn:bilayerdisordered_Hamiltonian}
  \nonumber
  \mathcal{H}_K \hspace{1mm} = && \hspace{1mm}
  \overbrace{- \hspace{1mm} t \sum_{<{\bf ij}>,\sigma}
  (a^\dagger_{{\bf i}\sigma} a_{{\bf j}\sigma} + h.c.)
  - t' \sum_{<\hspace{-1mm}<{\bf ii'}>\hspace{-1mm}>,\sigma}
  (a^\dagger_{{\bf i}\sigma} a_{{\bf i'}\sigma}
  + h.c.)}^{\mbox{\scriptsize{$A$-layer}}} \\
  \nonumber
  &&
  \overbrace{+ \hspace{1mm} t \sum_{<{\bf ij}>,\sigma}
  (b^\dagger_{{\bf i}\sigma} b_{{\bf j}\sigma} + h.c.)
  + t' \sum_{<\hspace{-1mm}<{\bf ii'}>\hspace{-1mm}>,\sigma}
  (b^\dagger_{{\bf i}\sigma} b_{{\bf i'}\sigma}
  + h.c.)}^{\mbox{\scriptsize{$B$-layer}}} \\
  \nonumber
  && \underbrace{- \hspace{1mm} \sum_{{\bf i},\sigma}
  t_h({\bf i}) (a^\dagger_{{\bf i}\sigma} b_{{\bf i}\sigma}
  + h.c.)}_{\mbox{\scriptsize{$A$-$B$ Layer hybridization}}}
  - \hspace{1mm} \mu \sum_{{\bf i},\sigma}
  (a^\dagger_{{\bf i}\sigma} a_{{\bf i}\sigma}
  + b^\dagger_{{\bf i}\sigma} b_{{\bf i}\sigma}); \\
  \nonumber
  && \underbrace{ + \hspace{1mm}
  \sum_{{\bf i}\in A,\sigma} V_d({\bf i})
  a^\dagger_{{\bf i}\sigma} a_{{\bf i}\sigma} +
  \sum_{{\bf i}\in B,\sigma} V_d({\bf i})
  b^\dagger_{{\bf i}\sigma} b_{{\bf i}\sigma}
  }_{Disorder \; term}; \\
  \mathcal{H}_U \hspace{1mm} = && \hspace{1mm}
  \underbrace{- \hspace{1mm} U \sum_{\bf i}
  (a^\dagger_{{\bf i}\uparrow} a^\dagger_{{\bf i}\downarrow}
  a_{{\bf i}\downarrow} a_{{\bf i}\uparrow}
  + b^\dagger_{{\bf i}\uparrow} b^\dagger_{{\bf i}\downarrow}
  b_{{\bf i}\downarrow} b_{{\bf i}\uparrow})
  }_{\mbox{\scriptsize{Interaction \; term}}}
\end{eqnarray}

  We recall that
$a^\dagger_{{\bf i}\sigma}$ ($b^\dagger_{{\bf i}\sigma}$) and
$a_{{\bf i}\sigma}$ ($b_{{\bf i}\sigma}$) are the creation and
annihilation operators of spin-$\frac{1}{2}$ fermions with spin
$\sigma = \uparrow, \downarrow$ at site ${\bf i}$ corresponding
to the $A$ layer ($B$ layer) of the bilayer square lattice. Here,
$t$ is the nearest-neighbor hopping, $t'$ is the next-nearest-neighbor
hopping, and $t_h$ is the interlayer hopping which hybridizes the
$A$ and $B$ layers, $U(>0)$ is the attractive
Hubbard interaction, and $\mu$ is the chemical potential. The
random potential $V_d^{A/B} ({\bf i})$ is chosen independently at each
site ${\bf i}$, belonging to layer $A$ or $B$, from the uniform
distribution $\bigl [-V_d:V_d \bigr ]$  that is symmetric about zero.
According to the central limit theorem,
$\sum_{{\bf i}=1}^N V_d({\bf i})=0$, where $N$ being the total
number of sites. The pure case corresponds to all on-site potentials
vanishing $(V_d({\bf i})=0)$.


\subsection{Brief Description of determinant quantum Monte Carlo Simulation}
\label{subsec:YPDDQMCDesc}

  We begin by including the disorder term of the Hamiltonian in
$\mathcal{H}_K$ and applying the \textit{Trotter-Suzuki}
decomposition to separate the kinetic and the interaction energy
exponentials \cite{Blankenbecler1981,Santos2003}. With the addition
of the disorder term in the kinetic energy (KE) term, the kinetic exponential
will have the following expression:
\begin{eqnarray}
  \label{eqn:KEVdexp}
  e^{-\Delta\tau \mathcal{\tilde{K}}} = \; &&
    \prod_\sigma e^{-\Delta\tau \sum_{\alpha,\gamma}
    \sum_{<{\bf ij}>} (c^\dagger_{{\bf i},\alpha,\sigma}
    \mathbb{\tilde{K}}^\sigma_{{\bf ij}{\alpha\gamma}}
     c_{{\bf j},\gamma,\sigma} + {\bf h.c.})}
\end{eqnarray}
where $c_\alpha$'s are equivalent to $a$ and $b$ operators for
$\alpha=1$ and $2$, respectively, and
$\mathbb{\tilde{K}}$ is the modified KE matrix
whose elements are given by
\begin{equation}
  \label{eqn:KijMat}
  \mathbb{\tilde{K}}^\sigma_{{\bf ij}{\alpha\gamma}} \; = \; 
         t_{{\bf ij}} \; - \; \bigl (\mu - V_d ({\bf i_\alpha})
         \bigr ) \; \delta_{{\bf ij}} \; \delta_{\alpha\gamma}
\end{equation}
with $t_{{\bf ij}}$ being the hopping matrix. The interaction exponential
has the following expression:
\begin{equation}
  \label{eqn:expV}
    e^{-\Delta\tau \mathcal{V}} = \;
    e^{-\Delta\tau \sum_\alpha \sum_{{\bf i}}
    U n^\alpha_{{\bf i}\uparrow}
              n^\alpha_{{\bf i}\downarrow}}
\end{equation}

  After applying the Hubbard-Stratonovich transformation for the
bilayer band-insulator model, the elements of the matrix $\mathbb{V}$
in the interaction exponential term are modified to
\begin{equation}
  \label{eqn:YPexpop}
    \mathbb{V}^\sigma_{{\bf ij}\alpha\gamma} \; = \;
    - \frac{\lambda s_{{\bf i}}}{\Delta\tau} \delta_{{\bf ij}}
     \; \delta_{\alpha \gamma}.
\end{equation}
with $cosh(\lambda) = e^{\abs{U}\Delta\tau/2}$ and $s_{{\bf i}}$ being the
discrete Ising variables, $s=\pm 1$. At half filling, $\mu$ is set as
$\mu = \frac{\abs{U}}{2}$. Following all the steps of the DQMC algorithm,
we performed the simulation for our model at half filling for $N=2\times L^2$
sites with the periodic boundary conditions. Here, $L$ represents the number
of sites in each direction of the square lattice. We choose the hopping
$t = 1$ to set our unit of energy. $t'/t = 0.1$ and $t_h/t = 0.6$ have been
set to compare the results with the ``clean" system studied in Refs.
\cite{Prasad2014,Prasad2022}. The inverse temperature has been discretized
in small imaginary time intervals $\Delta\tau \ t=0.05$, resulting in very small
systematic errors $(\sim \Delta\tau^2)$ involved in these simulations.
For $\abs{U}/t=8$, the $\Delta\tau \ t=0.025$ has been chosen to reduce
fluctuations in the high-interaction regime. All the simulations have been
done at temperature $T/t=0.1$ for a system size $L=16$, unless specified
otherwise. In all these calculations, disorder averages have been done
over $300-400$ disorder configurations, generated randomly from a uniform
distribution as discussed, and the indicated error bars are the statistical
error bars over these disorder averages of the Monte Carlo sampling.



\section{Results}
\label{sec:Results}



\subsection{Double Occupancy and the Pair Formation}

\begin{figure}[h]
  \centering
  \includegraphics[width=0.9\columnwidth]{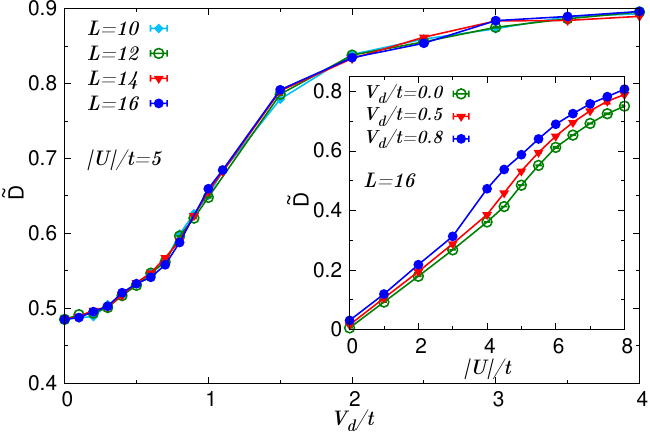}
  \caption{The evolution of the rescaled double occupancy
    $\tilde{D} = (4D-1)$ with random on-site disorder
    $V_d/t$ for various system sizes at the temperature
    $T/t = 0.1$. The system is at half filling, with
    interplane hybridization $t_h/t=0.6$. Inset: The plot
    of the $\tilde{D}$ vs the attractive interaction $\abs{U}/t$.}
  \label{fig:YPDdbleocc}
\end{figure}

  \Fig{fig:YPDdbleocc} shows the evolution of the rescaled
double occupancy $\tilde{D}$ or the density of the on-site pairs defined as
\begin{equation}
  \nonumber
  \tilde{D} = \frac{D - \mean{n_{{\bf i}\uparrow}}^2}
  {\mean{n_{{\bf i}\uparrow}} -
  \mean{n_{{\bf i}\uparrow}}^2} = 4 D-1,
\end{equation}
with the random on-site disorder $V_d/t$ for various system sizes.
Here the double occupancy $D = \frac{1}{N}\sum_{i,\alpha}\mean{n_{{\bf i}\uparrow}^\alpha
n_{{\bf i}\downarrow}^\alpha}$. The inset of \Fig{fig:YPDdbleocc} shows
the plot of the $\tilde{D}$ versus the attractive interaction $\abs{U}/t$
at temperature $T/t = 0.1$. We see that $\tilde{D}$ is independent of the
system size and it increases with the increase in the disorder strength
and saturates to its maximum value for the large disorder strengths. At
weak-disorder strengths $(V_d/t << 1)$, $\tilde{D}$ increases slowly as
the kinetic energy term dominates and favors delocalization, but in the
intermediate disorder region $(V_d/t \sim 1)$, the random disorder
potential competes with the KE term and hence enhances the pairing. As
we go towards the large disorder region, $\tilde{D}$ approaches its
limiting value and hence saturates. Thus the random on-site disorder promotes
$\tilde{D}$, the local pair formation, and hence the localization of
pairs, i.~e., the effect similar to the attractive interaction $\abs{U}/t$.
We see the existence of the molecule formation along the $BCS-BEC$
crossover as we tune the attractive interaction both in the absence
and in the presence of the random disorder which comes from the
evolution of the double occupancy (inset of \Fig{fig:YPDdbleocc}).
We see that the double occupancy increases from its noninteracting
limit value ($\tilde{D} \approx 0$) to its limiting value ($\tilde{D}
\approx 1$ at half filling) as $\abs{U}/t$ approaches infinity and the
presence of the disorder enhances this pair formation process.

\subsection{Effective Hopping}

\begin{figure}[h]
  \centerline{
  \includegraphics[width=0.9\columnwidth]{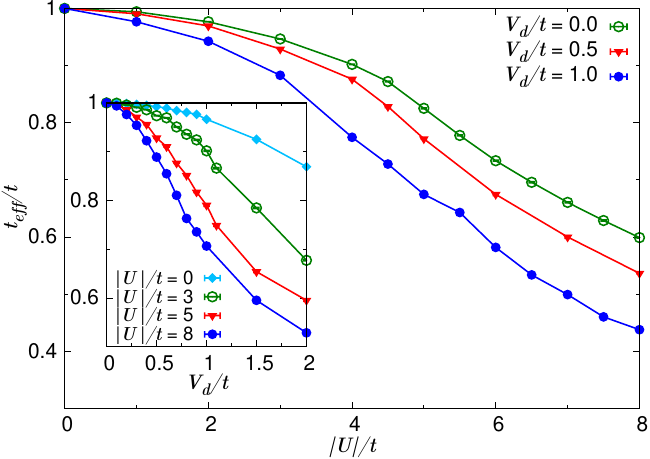}}
  \caption{The effective hopping
    $t_{eff}/t$ as a function of the interaction strength
    $\abs{U}/t$ for various disorder strengths $V_d/t$ at
    $T/t = 0.1$. As the interaction energy $\abs{U}/t$
    increases, the effective hopping declines. Inset:
    The effective hopping as a function $V_d/t$ for
    various $\abs{U}/t$ at the same temperature.}
  \label{fig:YPDKEeff}
\end{figure}

\begin{figure}[h]
  \centerline{
  \includegraphics[width=0.9\columnwidth]{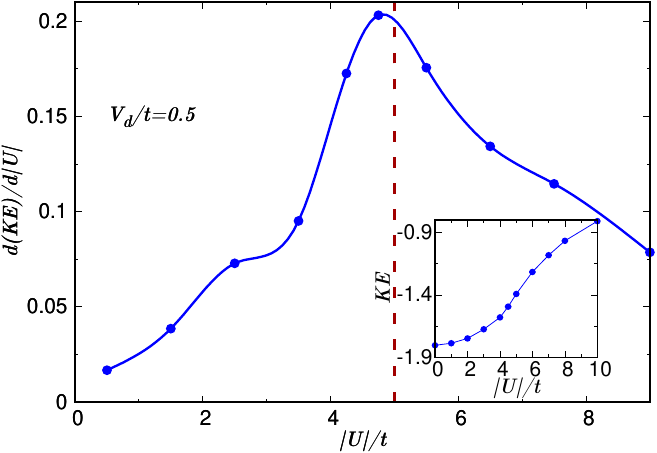}}
  \caption{The evolution of the kinetic energy (KE) and
    its derivative with respect to the attractive interaction
    $\abs{U}/t$ at disorder strength $V_d/t=0.5$. In
    the intermediate-coupling regime, there is a sharp increase
    in the KE which can be clearly seen in the
    derivative of KE, where we observe a peak at $\abs{U}/t=5$.}
  \label{fig:YPDKEvsU}
\end{figure}

  As we tune the disorder strength at finite attractive
interaction, we expect the effective single-fermion transfer to
decrease. Hence the local fermions tend to form pairs, resulting
in the increase in the double occupancy, as seen earlier. A
measure of this reduction in the single-fermion transfer is given
by the effective hopping defined as
\begin{equation}
  \label{eqn:YPDteff}
  \frac{t_{eff}}{t} =
  \frac {\mean{\mathcal{H}_K}}
  {\mean{\mathcal{H}_K}_{V=0}},
\end{equation}
the ratio of KE at finite disorder to the kinetic energy at zero
disorder at a given interaction strength $\abs{U}/t$. In
\Fig{fig:YPDKEeff}, we plot the effective hopping for various
disorder strengths $V_d/t$ at temperature $T/t=0.1$. We see that
the effective hopping declines as the attractive interaction
$\abs{U}/t$ increases and the declination gets faster as we
increase the disorder strength. Hence the disorder enhances the
pairing and reduces the effective hopping due to the localization
effects. We observe that the decrease in the effective hopping is
very sharp in the strong-coupling limit (inset of \Fig{fig:YPDKEeff})
in the presence of disorder, where the system goes to the Bose-glass
(BG) phase.

\begin{figure*}[ht]
  \centerline{
  \includegraphics[width=1.98\columnwidth]{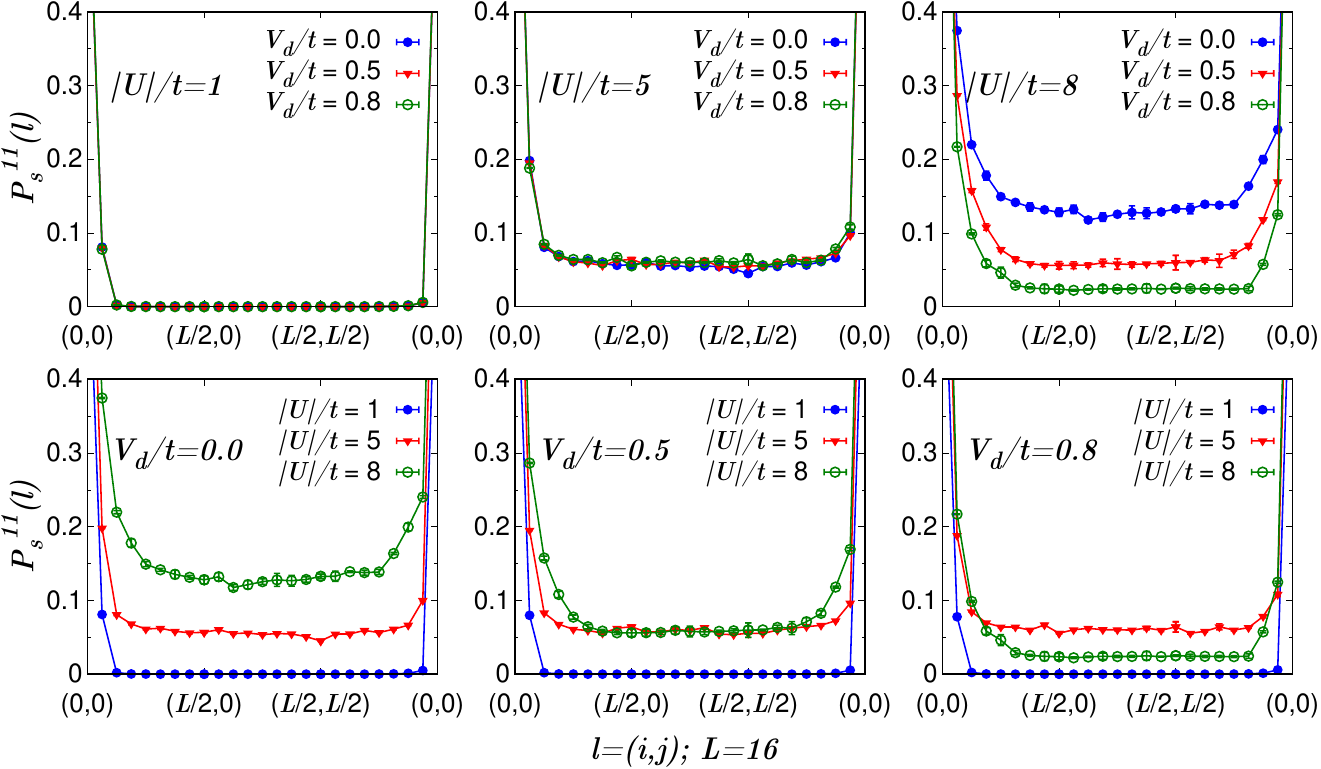}}
  \caption{The spatial dependence of the ground-state pair correlation
    functions $P_s^{11}(l)$ for different disorder strengths $V_d/t$
    in a bilayer BI model at attractive interaction $\abs{U}/t=1, 5$,
    and $8$ (top panels) and for different interaction strengths
    $\abs{U}/t$ at $V_d/t=0.0, 0.5$, and $0.8$ (bottom panels). 
    The correlation functions converge to a nonzero value at large
    separations for $\abs{U}/t=5$ and $8$, providing clear evidence
    for the long-range order even in the presence of random on-site
    disorder, though the value of the pair-pair correlations
    decreases with the increase in the disorder strength and goes to
    zero for $V_d/t=0.8$ at $\abs{U}/t=8$.}
  \label{fig:YPDpairlxly}
\end{figure*}

  In \Fig{fig:YPDKEvsU}, we have shown the evolution of KE and its
derivative with respect to the attractive interaction $\abs{U}/t$ at
disorder strength $V_d/t=0.5$. We observe a peak at $\abs{U}/t=5$ in
the derivative of KE, which coincides with the critical strength
$\abs{U_c}/t$, calculated from the finite-size scaling analysis of
the pair structure factor, which marks the transition from the BI to
the SF state. In the inset of \Fig{fig:YPDKEvsU}, we see that for the
interaction strengths $\abs{U}/t \sim 0-2$, the KE increases slowly,
but as we increase the interaction strength further, there is a sharp
rise in the KE (or a sharp decrease in the effective hopping).

\subsection{Pair-Pair Correlations}
\label{sec:DPC}

  In the following sections, we have studied the pair-pair and
density-density correlations in the presence of random on-site
disorder at half filling in the proposed bilayer band-insulator
model. We find that the pair-pair correlation survives in the
weak-disorder limit, while the density-density correlation
function gets suppressed even with a slight increase in the
disorder.

  \Fig{fig:YPDpairlxly} show the dependence of the ground state
pair-pair correlation functions, defined as
\begin{equation}
  \label{eqn:YPDPCF}
  P_s^{\alpha \gamma} ({\bf i},{\bf j}) =
  \mean{\Delta_s ({\bf i},\alpha)
  \Delta^\dagger_s ({\bf j},\gamma) + H.c.},
\end{equation}
on separation ${\bf i}$ for different combinations of the
disorder strengths $V_d$ and the attractive interactions
$\abs{U}/t$ in the bilayer BI model at half filling. Here,
$\Delta_s ({\bf i},\alpha) \sim c_{{\bf i},\alpha \downarrow}
c_{{\bf i},\alpha \uparrow}$ and $\Delta^\dagger_s ({\bf i},\alpha)$
are the pair annihilation and creation operators. As mentioned
earlier, the separation ${\bf l}=({\bf i},{\bf j})$ follows a
trajectory along the $x$ axis to maximal $x$ separation
$(\frac{L}{2},0)$ on a lattice with the periodic boundary
conditions, and then to $(\frac{L}{2},\frac{L}{2})$ before
returning to separation $(0,0)$. In the weak-coupling limit, there
is no pair-pair correlation (shown in \Fig{fig:YPDpairlxly} for
$\abs{U}/t = 1$) as the system remains in the BI. We see that the
correlation functions converge to a nonzero value at large
separations for $\abs{U}/t = 5$ in weak $(V_d/t << 1)$ and
intermediate $(V_d/t \sim 1)$ disorder regimes, providing a clear
evidence for the long-range order even in the presence of random
on-site disorder. At $\abs{U}/t = 8$, the pair correlation survives
in the weak-disorder limit, but goes to zero for $V_d/t = 0.8$,
indicating a transition from the SF to the BG phase where the fermionic
pairs get localized in the strong-coupling and the strong-disorder
limit. We observe that the pair correlations in the strong-coupling
regime gets strongly suppressed as compared to the intermediate-coupling
regimes. This reduces the SF region in the phase diagram. The existence
of the long-range order for $\abs{U}/t \geq 5$ implies that the
presence of the random on-site disorder does not change the
critical value of the interaction strength $\abs{U_c}/t$, which we
confirm from the finite-size scaling analysis.

\begin{figure}[h]
  \centerline{
  \includegraphics[width=0.9\columnwidth]{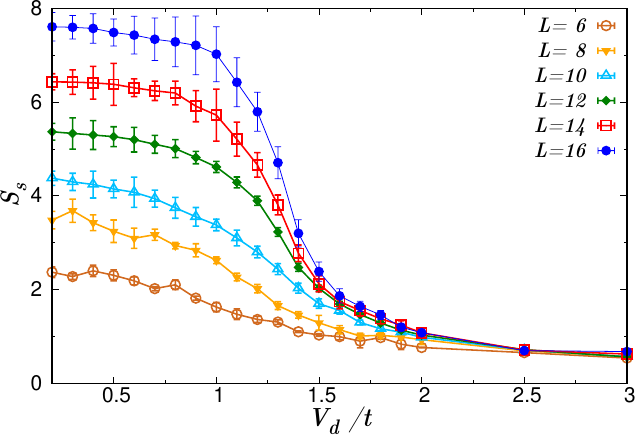}}
  \caption{The evolution of the $s$-wave pair structure factor $S_s$
    with the disorder strength $V_d/t$ for different system sizes at
    interaction $\abs{U}/t=5$. In the weak-disorder limit, $S_s$ almost
    remains constant. With the increase in the disorder strength, $S_s$
    slightly increases and then decreases sharply, finally saturating
    to a finite nonzero value which does not depend on the size of the
    lattice. $S_s$ depends on the system size in the weak-disorder limit.}
  \label{fig:YPDPsvsV}
\end{figure}

  In \Fig{fig:YPDPsvsV}, we show the evolution of the $s$-wave
pair structure factor $S_s$ with the disorder strength $V_d/t$ for
various system sizes for interaction $\abs{U}/t = 5$ at
$T/t = 0.1$. We observe that in the weak-disorder limit, the pair
structure factor increases slightly from its ``clean" system
(absence of disorder) value and then decreases sharply with the
increase in the disorder strength, finally saturating to a finite
nonzero value which does not depend on the size of the lattice.
It shows that the $s$-wave pair structure factor has a strong
system-size dependence in the weak-disorder limit, indicating that
the correlation length $\xi$, which depends on the disorder
strength and temperature, is large compared to the lattice
size $L$. The system-size dependence goes away as soon as $\xi$
becomes small as compared to $L$, which gives the information
about the short-range nature of the pair-pair correlation in the
large disorder limit.

\subsection{Scaling analysis}
\label{sec:DScaling}

\begin{figure}[h]
  \centerline{
  \includegraphics[width=0.98\columnwidth]{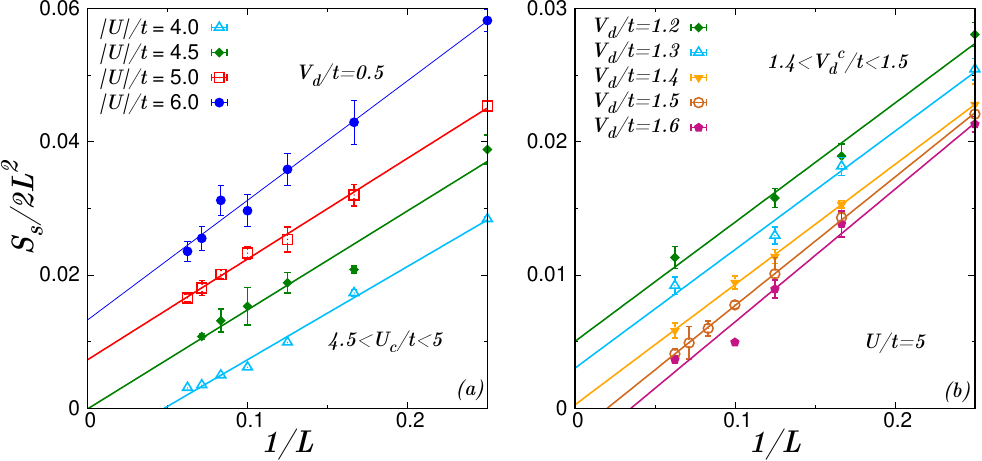}}
  \caption{Finite-size scaling of the $s$-wave pair structure factor $S_s/2/L^2$.
    The symbols are the determinant quantum Monte Carlo results and the dashed
    lines are the extrapolation performed via a linear least-squares fit for (a)
    each $\abs{U}/t$ in presence of disorder $V_d/t=0.5$ and (b) each
    $V_d/t$ at $\abs{U}/t=5$. The inverse temperature has been fixed at
    $\beta\ t=10$. We observe that $S_s$ vanishes for $\abs{U}/t<4.5$
    as $L\rightarrow\infty$. Thus, $\abs{U_c}/t$ lies in the range
    $4.5<\abs{U_c}/t<5$.}
  \label{fig:YPDPsvsL}
\end{figure}

  We have observed that our bilayer BI model displays
the long-range order in the pair-pair correlations, even at
intermediate disorder strengths. Hence, we expect the Huse's
argument \cite{Huse1988} of the ``spin-wave scaling'' to hold,
\begin{equation}
  \label{eqn:PsDDelta}
  \frac{S_s}{2L^2} = \varDelta_0^2|_V + \frac{C(U,V_d)}{L}
\end{equation}
where $\varDelta_0$ is the SF order parameter at zero temperature
and disorder $V_d/t$, and $C$ is a constant which depends on the
interaction strength $U/t$ and random on-site disorder $V_d/t$.

The SF order parameter $\varDelta_0$ can also be extracted
from the {\it equal-time} $s$-wave pair-pair correlation function
$P_s ({\bf l})$ for the two most distant points on a lattice, i.e.,
having ${\bf R} = (L/2,L/2)$
\cite{Scalettar1999}, with a similar spin-wave theory correction,
\vspace{-2mm}
\begin{equation}
  \label{eqn:PairDelta}
  P_s ({\bf R}) = \varDelta_0^2 + B(U,V_d) L.
\vspace{-2mm}
\end{equation}
Thus, we estimate the zero-temperature SF parameter from the finite-size
scaling of the $s$-wave pair structure factor using \Eqn{eqn:PsDDelta} and
\Eqn{eqn:PairDelta}, and hence estimate the zero-temperature
critical value of the interaction at which our bilayer BI, in the presence
of disorder, undergoes a transition to the SF state.


In \Fig{fig:YPDPsvsL}, we present the finite-size scaling of the
$s$-wave pair structure factor $S_s/2L^2$ in the presence of the random
disorder. It shows that the zero-temperature critical interaction
$\abs{U}/t$ lies between $4.5$ and $5.0$ at the disorder strength
$V_d/t = 0.5$, which is the same as is obtained in the absence of
disorder in Ref. \cite{Prasad2022}. Hence the disorder does not affect
the critical value, but plays a significant role in suppressing
the pair correlation function.

\begin{figure}[h]
  \centerline{
  \includegraphics[width=0.9\columnwidth]{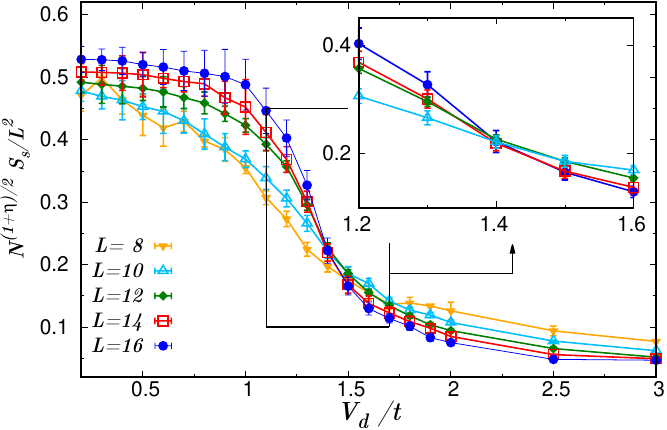}}
  \caption{Rescaled $S_s$ as a function of the disorder strength
    $V_d/t$ at $\abs{U}/t=5$ for different system sizes. Inset:
    The enlarged region where the curves intercept each other,
    around $V_d/t=1.5-2$.}
  \label{fig:YPDScaledPsvsV}
\end{figure}

\begin{figure}[h]
  \centerline{
  \includegraphics[width=0.9\columnwidth]{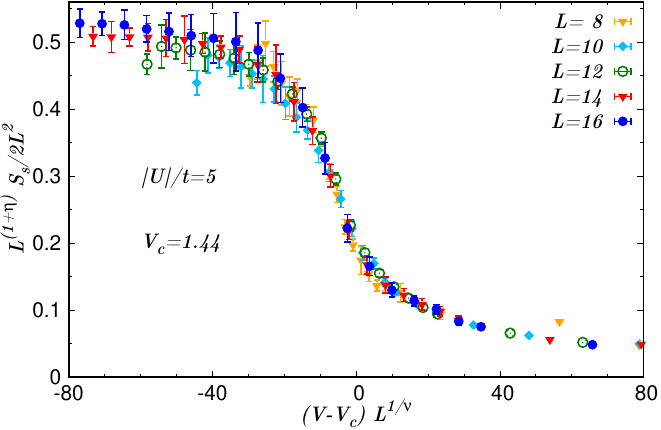}}
  \caption{Rescaled $S_s$ plotted against $\abs{V_d-V_c^d} \
    L^{1/\nu}$ at $\abs{U}/t=5$ for different system sizes. All the
    data points of the various system sizes collapse into a single
    curve for $\nu=0.67$,$\eta=0.04$ and $V_d^c/t=1.44$.}
  \label{fig:YPDPsfit}
\end{figure}

  In the strong-coupling limit, due to a large on-site attraction,
the fermions form tightly bound pairs and can be treated as
bosons which condense to form a SF phase. In this limit,
we can map our attractive Hubbard model to the effective
hard-core Bose-Hubbard model with repulsive next-nearest-neighbor
interaction. The pair annihilation [$\Delta_s ({\bf i},\alpha) \sim
c_{{\bf i},\alpha \downarrow} c_{{\bf i},\alpha \uparrow}$] and
creation [$\Delta^\dagger_s ({\bf i},\alpha)$] operators of our
proposed model will be equivalent to the bosonic creation and
annihilation operators $(b^\dagger_{{\bf i}}$ and $b_{{\bf i}})$.
Thus we expect that in the presence of disorder, there exists a BG
phase before the system goes to a CDW insulator state since, at
half filling, in the strong-coupling limit, a hard-core Bose-Hubbard
model exhibits a SF to BG transition with increasing disorder $V_d/t$.
This transition belongs to the $(d+1)-XY$ universality class
\cite{Fisher1989}.


  To estimate the critical value of the disorder strength beyond
which our system undergoes a SF to BG transition,
we use the scaling ansatz used in Refs. \cite{Fisher1989,Mondaini2015},
\begin{equation}
  \label{eqn:YPDPairScaling}
  L^{1+\eta} \; \frac{S_s}{L^2} = F\bigl ((V_d-V_c) \;
  L^{1/\nu} \bigr )
\end{equation}
where $\nu$ and $\eta$ are the correlation length exponent and the
order parameter exponent, respectively. $V_c/t$ is the critical
disorder strength required to destroy the superfluid order. At
$V_d/t = V_c/t$, the rescaled pair structure factor becomes
independent of the system size and hence all the curves for
different system sizes must intercept each other at $V_c/t$.

  \Fig{fig:YPDScaledPsvsV} shows the rescaled pair structure factor
$S_s$ as a function of the disorder strength $V_d/t$ at $\abs{U}/t=5$
for different system sizes. We observe that all the curves
corresponding to different system sizes intersect each other at
$V_d^c/t\sim 1.4$. The inset shows the enlarged region, around
$V_d/t = 1.1-1.7$, where the curves intersect each other. In
\Fig{fig:YPDPsfit}, we plot the rescaled pair structure factor $S_s$
versus the universal scaling function $F(z)$ [\Eqn{eqn:YPDPairScaling}].
We observe that all the curves corresponding to different system sizes
collapse to a single curve for $\nu = 0.67$, $\eta = 0.04$, and
$V_d^c/t = 1.44$, except in the weak-disorder regime. The perfect
collapse of our data, for $\nu = 0.67$ and $\eta = 0.04$, shows
that the SF to BG transition lies in the universality class of the
$3D-XY$ model. The obtained critical disorder strength is roughly the
same as the single-layer half-filled attractive Hubbard model
where $V_d^c/t\sim 1.5$ at $\abs{U}/t=4$ \cite{Huscroft1997}.


\begin{figure}[h]
  \centerline{
  \includegraphics[width=0.98\columnwidth]{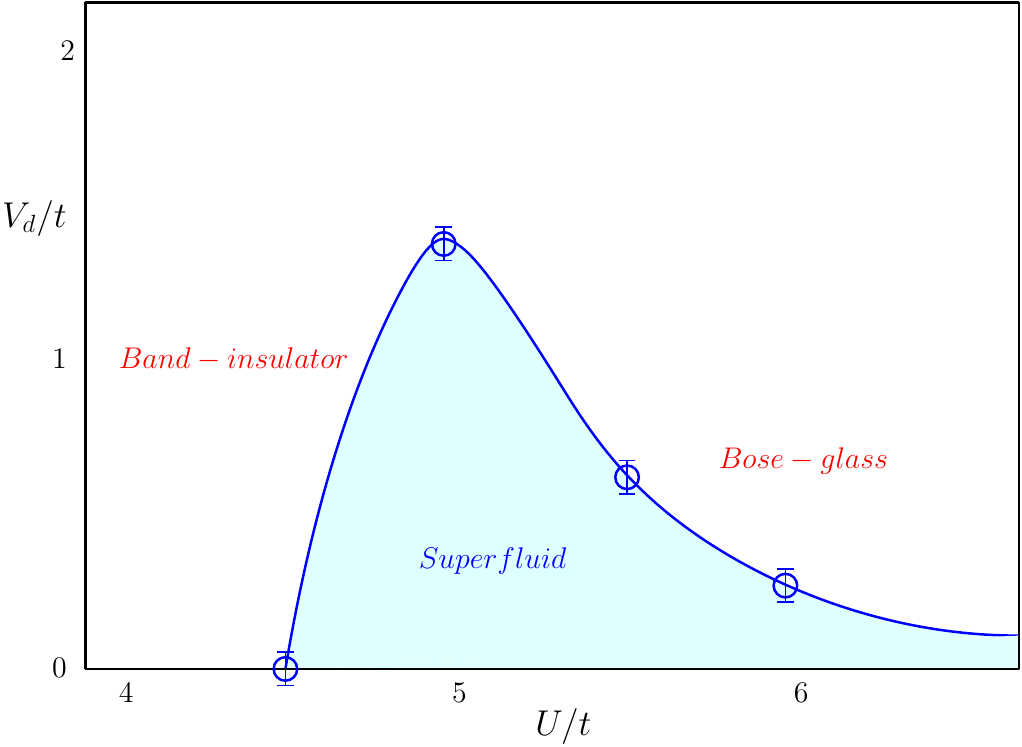}}
  \caption{$V-U$ phase diagram of proposed bilayer band insulator
    (BI) in the presence of disorder at finite layer hybridization.
    In the absence of disorder, the system goes to a superfluid (SF) phase
    from the BI phase as we tune the interaction. In the strong-coupling
    limit, it goes to a ``bosonic" charge density wave (CDW) phase. As
    we tune the disorder, the BI phase is expected to go to an
    Anderson-insulating phase in the large disorder limit. We observe that
    weak disorder suppresses the SF phase largely in the strong-coupling limit.
    Hence the region of the SF phase reduces in the presence of random
    on-site disorder. In the strong-coupling limit, based on the mapping
    to the hard-core Bose-Hubbard model, the system is expected to go from
    the SF phase to the BG phase. Blue circles denote the critical
    disorder determined by the scaling analysis for $U/t = 5, 5.5$, and $6$,
    respectively.}
  \label{fig:YPDPhaseDiag}
\end{figure}

\subsection{$V-U$ phase diagram of bilayer band insusulator in the
  presence of disorder.}
\label{sec:DScaling}

Finally, we discuss a $V-U$ phase diagram for the proposed bilayer BI model
at finite hopping between the layers and in the presence of disorder, shown
in \Fig{fig:YPDPhaseDiag}. In the ``clean'' noninteracting case, as we tune
disorder, we expect that the BI will eventually go to the Anderson-insulator
state for higher values of the disorder strengths. Beyond the critical
interaction, tuning of the disorder is expected to suppress the pair-pair
correlations. In the strong-coupling limit $(\abs{U} >> t)$, the fermions
exist in a bound state and hence the system can be described by a hard-core
bosonic Hubbard model with repulsive next-nearest-neighbor interactions. As we
expect that in the strong coupling limit at noncommensurate integer filling
$(n=0.5)$, the hard-core Bose-Hubbard model shows a SF to BG transition
\cite{Fisher1989}. Based on this mapping, we can expect our system to go
from a SF to BG phase in the presence of disorder. We observe that weak disorder
suppresses the SF phase largely in the strong-coupling limit, leading to the
reduction of the SF region in the presence of random on-site disorder.

\begin{table}
\begin{ruledtabular}
\begin{tabular}{p{0.5in} p{1.25in} p{1.25in}}
  \rowcolor{gray!10}
  {\centering System} & {\centering Single layer model} & {\centering Bilayer model}\\
  \hline
  \hline
  {\centering \cellcolor {gray!10}$V=0$}   & Degenerate SF+CDW & BI to SF transition\\
  {\centering \cellcolor {gray!10}$V\ne0$} & SF to BG transition & BI to SF transition; SF to BG transition \\
\end{tabular}
\end{ruledtabular}
\caption{Comparison of presented bilayer band-insulator model with single-layer attractive
  Hubbard model{\label{TB:Single_Bilayer_Comp}}}
\end{table}

\section{Concluding Remarks}
\label{sec:Concl}

  In this paper, we have studied the bilayer BI model in the
presence of disorder. Using the DQMC numerical technique, we have shown
the effect of disorder on the KE and the double occupancy. We observe
that the on-site random disorder plays a significant role in the
localization of on-site pairs, and hence in the reduction of the effective
hopping. This results in an increase in the double occupancy, which is
an effect similar to the attractive interaction.

  We also observe the existence of the long-range order in the
pair-pair correlations at various disorder strengths. The random
disorder does not affect the critical value of the interaction
strength and the SF state survives even at intermediate
disorder strengths. Finally, via finite-size scaling analysis,
we have computed the critical disorder strength $V_c\sim 1.44$.
For a single-layer attractive Hubbard model on a square lattice,
previous QMC studies \cite{Huscroft1997} reported $V_c \sim 1.5$
at $\abs{U}/t=4$. In \Tabs{TB:Single_Bilayer_Comp}, we have summarized
the differences between the single-layer attractive Hubbard model
with the bilayer band-insulator model studied here. In the single-layer
AHM, the ground state is degenerate with the superfluid along with the
CDW state for any finite interaction in the absence of disorder. Any
finite amount of disorder removes this degeneracy with only the SF state
in the ground state. At $V_c$, we observe the SF to BG transition. In
the bilayer band-insulating model presented here, we have a BI to SF
transition at finite interaction $U_c$ in the absence of any disorder
and there is no leading CDW order up to $U=10$. In the presence of
disorder, there is transition from a BI to SF phase and from SF to BG
phase in the ground state.

Note that for large energy scale parameters $t$, $\beta$, and $U$,
small $\Delta\tau$ is necessary for the accuracy of the Trotter-Suzuki
decomposition \cite{Bai2009}. Generally, the imaginary time steps are
chosen such that $\Delta \tau \le \sqrt{0.25/U}$. As $U$ increases, one
further need to decrease the imaginary time step, which further increases
the running time of programs. The finiteness of $\Delta \tau$ is a source
of systematic errors. In this work, for practical purposes, the inverse
temperature has been discretized in $M=200$ small imaginary time intervals
$\Delta \tau t = 0.05$, which shows large fluctuations in two-particle
correlations at large interaction regime (though disorder seems to destroy
these fluctuations). For $|U|/t=8$, we reduced the size of the imaginary
time step to $\Delta \tau t = 0.025$ by increasing the number of grids,
$M=400$.




\section*{Acknowledgements}

This work was supported by Ministry of Science, Republic of Korea through Grant No.
NRF-2021R1111A2057259. Y.P. would like to acknowledge CSIR for financial support. Y.P.
thanks A. V. Mallik, A. Halder, V. B. Shenoy, and Nandini Trivedi for various
discussions and comments. Y.P. would also like to thank V. B. Shenoy
for the cluster usage. H.L. acknowledges the hospitality at APCTP where part of
this work was done.


\bibliography{Bibliography.bib}

\begin{thebibliography}{30}%
\makeatletter
\providecommand \@ifxundefined [1]{%
 \@ifx{#1\undefined}
}%
\providecommand \@ifnum [1]{%
 \ifnum #1\expandafter \@firstoftwo
 \else \expandafter \@secondoftwo
 \fi
}%
\providecommand \@ifx [1]{%
 \ifx #1\expandafter \@firstoftwo
 \else \expandafter \@secondoftwo
 \fi
}%
\providecommand \natexlab [1]{#1}%
\providecommand \enquote  [1]{``#1''}%
\providecommand \bibnamefont  [1]{#1}%
\providecommand \bibfnamefont [1]{#1}%
\providecommand \citenamefont [1]{#1}%
\providecommand \href@noop [0]{\@secondoftwo}%
\providecommand \href [0]{\begingroup \@sanitize@url \@href}%
\providecommand \@href[1]{\@@startlink{#1}\@@href}%
\providecommand \@@href[1]{\endgroup#1\@@endlink}%
\providecommand \@sanitize@url [0]{\catcode `\\12\catcode `\$12\catcode
  `\&12\catcode `\#12\catcode `\^12\catcode `\_12\catcode `\%12\relax}%
\providecommand \@@startlink[1]{}%
\providecommand \@@endlink[0]{}%
\providecommand \url  [0]{\begingroup\@sanitize@url \@url }%
\providecommand \@url [1]{\endgroup\@href {#1}{\urlprefix }}%
\providecommand \urlprefix  [0]{URL }%
\providecommand \Eprint [0]{\href }%
\providecommand \doibase [0]{https://doi.org/}%
\providecommand \selectlanguage [0]{\@gobble}%
\providecommand \bibinfo  [0]{\@secondoftwo}%
\providecommand \bibfield  [0]{\@secondoftwo}%
\providecommand \translation [1]{[#1]}%
\providecommand \BibitemOpen [0]{}%
\providecommand \bibitemStop [0]{}%
\providecommand \bibitemNoStop [0]{.\EOS\space}%
\providecommand \EOS [0]{\spacefactor3000\relax}%
\providecommand \BibitemShut  [1]{\csname bibitem#1\endcsname}%
\let\auto@bib@innerbib\@empty
\bibitem [{\citenamefont {Anderson}(1959)}]{Anderson1959}%
  \BibitemOpen
  \bibfield  {author} {\bibinfo {author} {\bibfnamefont {P.~W.}\ \bibnamefont
  {Anderson}},\ }\bibfield  {title} {\bibinfo {title} {Theory of dirty
  superconductors},\ }\href
  {https://doi.org/http://dx.doi.org/10.1016/0022-3697(59)90036-8} {\bibfield
  {journal} {\bibinfo  {journal} {Journal of Physics and Chemistry of Solids}\
  }\textbf {\bibinfo {volume} {11}},\ \bibinfo {pages} {26} (\bibinfo {year}
  {1959})}\BibitemShut {NoStop}%
\bibitem [{\citenamefont {Trivedi}\ \emph {et~al.}(1996)\citenamefont
  {Trivedi}, \citenamefont {Scalettar},\ and\ \citenamefont
  {Randeria}}]{Trivedi1996}%
  \BibitemOpen
  \bibfield  {author} {\bibinfo {author} {\bibfnamefont {N.}~\bibnamefont
  {Trivedi}}, \bibinfo {author} {\bibfnamefont {R.~T.}\ \bibnamefont
  {Scalettar}},\ and\ \bibinfo {author} {\bibfnamefont {M.}~\bibnamefont
  {Randeria}},\ }\bibfield  {title} {\bibinfo {title} {Superconductor-insulator
  transition in a disordered electronic system},\ }\href
  {https://doi.org/10.1103/PhysRevB.54.R3756} {\bibfield  {journal} {\bibinfo
  {journal} {Physical Review B}\ }\textbf {\bibinfo {volume} {54}},\ \bibinfo
  {pages} {R3756} (\bibinfo {year} {1996})}\BibitemShut {NoStop}%
\bibitem [{\citenamefont {Scalettar}\ \emph {et~al.}(1999)\citenamefont
  {Scalettar}, \citenamefont {Trivedi},\ and\ \citenamefont
  {Huscroft}}]{Scalettar1999}%
  \BibitemOpen
  \bibfield  {author} {\bibinfo {author} {\bibfnamefont {R.~T.}\ \bibnamefont
  {Scalettar}}, \bibinfo {author} {\bibfnamefont {N.}~\bibnamefont {Trivedi}},\
  and\ \bibinfo {author} {\bibfnamefont {C.}~\bibnamefont {Huscroft}},\
  }\bibfield  {title} {\bibinfo {title} {Quantum {Monte Carlo} study of the
  disordered attractive {Hubbard} model},\ }\href
  {https://doi.org/10.1103/PhysRevB.59.4364} {\bibfield  {journal} {\bibinfo
  {journal} {Physical Review B}\ }\textbf {\bibinfo {volume} {59}},\ \bibinfo
  {pages} {4364} (\bibinfo {year} {1999})}\BibitemShut {NoStop}%
\bibitem [{\citenamefont {Dynes}\ \emph {et~al.}(1984)\citenamefont {Dynes},
  \citenamefont {Garno}, \citenamefont {Hertel},\ and\ \citenamefont
  {Orlando}}]{Dynes1984}%
  \BibitemOpen
  \bibfield  {author} {\bibinfo {author} {\bibfnamefont {R.~C.}\ \bibnamefont
  {Dynes}}, \bibinfo {author} {\bibfnamefont {J.~P.}\ \bibnamefont {Garno}},
  \bibinfo {author} {\bibfnamefont {G.~B.}\ \bibnamefont {Hertel}},\ and\
  \bibinfo {author} {\bibfnamefont {T.~P.}\ \bibnamefont {Orlando}},\
  }\bibfield  {title} {\bibinfo {title} {Tunneling study of superconductivity
  near the metal--insulator transition},\ }\href
  {https://doi.org/10.1103/PhysRevLett.53.2437} {\bibfield  {journal} {\bibinfo
   {journal} {Physical Review Letters}\ }\textbf {\bibinfo {volume} {53}},\
  \bibinfo {pages} {2437} (\bibinfo {year} {1984})}\BibitemShut {NoStop}%
\bibitem [{\citenamefont {White}\ \emph {et~al.}(1986)\citenamefont {White},
  \citenamefont {Dynes},\ and\ \citenamefont {Garno}}]{White1986}%
  \BibitemOpen
  \bibfield  {author} {\bibinfo {author} {\bibfnamefont {A.~E.}\ \bibnamefont
  {White}}, \bibinfo {author} {\bibfnamefont {R.~C.}\ \bibnamefont {Dynes}},\
  and\ \bibinfo {author} {\bibfnamefont {J.~P.}\ \bibnamefont {Garno}},\
  }\bibfield  {title} {\bibinfo {title} {Destruction of superconductivity in
  quench--condensed two--dimensional films},\ }\href
  {https://doi.org/10.1103/PhysRevB.33.3549} {\bibfield  {journal} {\bibinfo
  {journal} {Physical Review B}\ }\textbf {\bibinfo {volume} {33}},\ \bibinfo
  {pages} {3549} (\bibinfo {year} {1986})}\BibitemShut {NoStop}%
\bibitem [{\citenamefont {Abrikosov}\ and\ \citenamefont
  {Gor\'kov}(1959)}]{Abrikosov1959}%
  \BibitemOpen
  \bibfield  {author} {\bibinfo {author} {\bibfnamefont {A.~A.}\ \bibnamefont
  {Abrikosov}}\ and\ \bibinfo {author} {\bibfnamefont {L.~P.}\ \bibnamefont
  {Gor\'kov}},\ }\bibfield  {title} {\bibinfo {title} {Superconducting alloys
  at finite temperatures},\ }\href
  {http://www.jetp.ac.ru/cgi-bin/dn/e_009_01_0220.pdf} {\bibfield  {journal}
  {\bibinfo  {journal} {Sov. Phys. J. Expt and Theor. Phys.}\ }\textbf
  {\bibinfo {volume} {36}},\ \bibinfo {pages} {319} (\bibinfo {year}
  {1959})}\BibitemShut {NoStop}%
\bibitem [{\citenamefont {Ma}\ and\ \citenamefont {Lee}(1985)}]{Ma1985}%
  \BibitemOpen
  \bibfield  {author} {\bibinfo {author} {\bibfnamefont {M.}~\bibnamefont
  {Ma}}\ and\ \bibinfo {author} {\bibfnamefont {P.~A.}\ \bibnamefont {Lee}},\
  }\bibfield  {title} {\bibinfo {title} {Localized superconductors},\ }\href
  {https://doi.org/10.1103/PhysRevB.32.5658} {\bibfield  {journal} {\bibinfo
  {journal} {Physical Review B}\ }\textbf {\bibinfo {volume} {32}},\ \bibinfo
  {pages} {5658} (\bibinfo {year} {1985})}\BibitemShut {NoStop}%
\bibitem [{\citenamefont {Orignac}\ and\ \citenamefont
  {Giamarchi}(1996)}]{Orignac1996}%
  \BibitemOpen
  \bibfield  {author} {\bibinfo {author} {\bibfnamefont {E.}~\bibnamefont
  {Orignac}}\ and\ \bibinfo {author} {\bibfnamefont {T.}~\bibnamefont
  {Giamarchi}},\ }\bibfield  {title} {\bibinfo {title} {Effects of weak
  disorder on two coupled {Hubbard} chains},\ }\href
  {https://doi.org/10.1103/PhysRevB.53.R10453} {\bibfield  {journal} {\bibinfo
  {journal} {Phys. Rev. B}\ }\textbf {\bibinfo {volume} {53}},\ \bibinfo
  {pages} {R10453} (\bibinfo {year} {1996})}\BibitemShut {NoStop}%
\bibitem [{\citenamefont {Paiva}\ \emph {et~al.}(2015)\citenamefont {Paiva},
  \citenamefont {Khatami}, \citenamefont {Yang}, \citenamefont {Rousseau},
  \citenamefont {Jarrell}, \citenamefont {Moreno}, \citenamefont {Hulet},\ and\
  \citenamefont {Scalettar}}]{Paiva2015}%
  \BibitemOpen
  \bibfield  {author} {\bibinfo {author} {\bibfnamefont {T.}~\bibnamefont
  {Paiva}}, \bibinfo {author} {\bibfnamefont {E.}~\bibnamefont {Khatami}},
  \bibinfo {author} {\bibfnamefont {S.}~\bibnamefont {Yang}}, \bibinfo {author}
  {\bibfnamefont {V.}~\bibnamefont {Rousseau}}, \bibinfo {author}
  {\bibfnamefont {M.}~\bibnamefont {Jarrell}}, \bibinfo {author} {\bibfnamefont
  {J.}~\bibnamefont {Moreno}}, \bibinfo {author} {\bibfnamefont {R.~G.}\
  \bibnamefont {Hulet}},\ and\ \bibinfo {author} {\bibfnamefont {R.~T.}\
  \bibnamefont {Scalettar}},\ }\bibfield  {title} {\bibinfo {title} {Cooling
  atomic gases with disorder},\ }\href
  {https://doi.org/10.1103/PhysRevLett.115.240402} {\bibfield  {journal}
  {\bibinfo  {journal} {Physical Review Letters}\ }\textbf {\bibinfo {volume}
  {115}},\ \bibinfo {pages} {240402} (\bibinfo {year} {2015})}\BibitemShut
  {NoStop}%
\bibitem [{\citenamefont {Damski}\ \emph {et~al.}(2003)\citenamefont {Damski},
  \citenamefont {Zakrzewski}, \citenamefont {Santos}, \citenamefont {Zoller},\
  and\ \citenamefont {Lewenstein}}]{Damski2003}%
  \BibitemOpen
  \bibfield  {author} {\bibinfo {author} {\bibfnamefont {B.}~\bibnamefont
  {Damski}}, \bibinfo {author} {\bibfnamefont {J.}~\bibnamefont {Zakrzewski}},
  \bibinfo {author} {\bibfnamefont {L.}~\bibnamefont {Santos}}, \bibinfo
  {author} {\bibfnamefont {P.}~\bibnamefont {Zoller}},\ and\ \bibinfo {author}
  {\bibfnamefont {M.}~\bibnamefont {Lewenstein}},\ }\bibfield  {title}
  {\bibinfo {title} {Atomic {Bose} and {Anderson} glasses in optical
  lattices},\ }\href {https://doi.org/10.1103/PhysRevLett.91.080403} {\bibfield
   {journal} {\bibinfo  {journal} {Physical Review Letters}\ }\textbf {\bibinfo
  {volume} {91}},\ \bibinfo {pages} {080403} (\bibinfo {year}
  {2003})}\BibitemShut {NoStop}%
\bibitem [{\citenamefont {Lye}\ \emph {et~al.}(2005)\citenamefont {Lye},
  \citenamefont {Fallani}, \citenamefont {Modugno}, \citenamefont {Wiersma},
  \citenamefont {Fort},\ and\ \citenamefont {Inguscio}}]{Lye2005}%
  \BibitemOpen
  \bibfield  {author} {\bibinfo {author} {\bibfnamefont {J.~E.}\ \bibnamefont
  {Lye}}, \bibinfo {author} {\bibfnamefont {L.}~\bibnamefont {Fallani}},
  \bibinfo {author} {\bibfnamefont {M.}~\bibnamefont {Modugno}}, \bibinfo
  {author} {\bibfnamefont {D.~S.}\ \bibnamefont {Wiersma}}, \bibinfo {author}
  {\bibfnamefont {C.}~\bibnamefont {Fort}},\ and\ \bibinfo {author}
  {\bibfnamefont {M.}~\bibnamefont {Inguscio}},\ }\bibfield  {title} {\bibinfo
  {title} {{Bose--Einstein} condensate in a random potential},\ }\href
  {https://doi.org/10.1103/PhysRevLett.95.070401} {\bibfield  {journal}
  {\bibinfo  {journal} {Physical Review Letters}\ }\textbf {\bibinfo {volume}
  {95}},\ \bibinfo {pages} {070401} (\bibinfo {year} {2005})}\BibitemShut
  {NoStop}%
\bibitem [{\citenamefont {Cl\'ement}\ \emph {et~al.}(2006)\citenamefont
  {Cl\'ement}, \citenamefont {Var\'on}, \citenamefont {Retter}, \citenamefont
  {Sanchez-Palencia}, \citenamefont {Aspect},\ and\ \citenamefont
  {Bouyer}}]{Clement2006}%
  \BibitemOpen
  \bibfield  {author} {\bibinfo {author} {\bibfnamefont {D.}~\bibnamefont
  {Cl\'ement}}, \bibinfo {author} {\bibfnamefont {A.~F.}\ \bibnamefont
  {Var\'on}}, \bibinfo {author} {\bibfnamefont {J.~A.}\ \bibnamefont {Retter}},
  \bibinfo {author} {\bibfnamefont {L.}~\bibnamefont {Sanchez-Palencia}},
  \bibinfo {author} {\bibfnamefont {A.}~\bibnamefont {Aspect}},\ and\ \bibinfo
  {author} {\bibfnamefont {P.}~\bibnamefont {Bouyer}},\ }\bibfield  {title}
  {\bibinfo {title} {Experimental study of the transport of coherent
  interacting matter--waves in a {1D} random potential induced by laser
  speckle},\ }\href {https://doi.org/10.1088/1367-2630/8/8/165} {\bibfield
  {journal} {\bibinfo  {journal} {New Journal of Physics}\ }\textbf {\bibinfo
  {volume} {8}},\ \bibinfo {pages} {010165} (\bibinfo {year}
  {2006})}\BibitemShut {NoStop}%
\bibitem [{\citenamefont {Pasienski}\ \emph {et~al.}(2010)\citenamefont
  {Pasienski}, \citenamefont {McKay}, \citenamefont {White},\ and\
  \citenamefont {DeMarco}}]{Pasienski2010}%
  \BibitemOpen
  \bibfield  {author} {\bibinfo {author} {\bibfnamefont {M.}~\bibnamefont
  {Pasienski}}, \bibinfo {author} {\bibfnamefont {D.}~\bibnamefont {McKay}},
  \bibinfo {author} {\bibfnamefont {M.}~\bibnamefont {White}},\ and\ \bibinfo
  {author} {\bibfnamefont {B.}~\bibnamefont {DeMarco}},\ }\bibfield  {title}
  {\bibinfo {title} {A disordered insulator in an optical lattice},\ }\href
  {https://doi.org/10.1038/nphys1726} {\bibfield  {journal} {\bibinfo
  {journal} {Nature Physics}\ }\textbf {\bibinfo {volume} {6}},\ \bibinfo
  {pages} {677} (\bibinfo {year} {2010})}\BibitemShut {NoStop}%
\bibitem [{\citenamefont {Billy}\ \emph {et~al.}(2008)\citenamefont {Billy},
  \citenamefont {Josse}, \citenamefont {Zuo}, \citenamefont {Bernard},
  \citenamefont {Hambrecht}, \citenamefont {Lugan}, \citenamefont {Clement},
  \citenamefont {Sanchez-Palencia}, \citenamefont {Bouyer},\ and\ \citenamefont
  {Aspect}}]{Billy2008}%
  \BibitemOpen
  \bibfield  {author} {\bibinfo {author} {\bibfnamefont {J.}~\bibnamefont
  {Billy}}, \bibinfo {author} {\bibfnamefont {V.}~\bibnamefont {Josse}},
  \bibinfo {author} {\bibfnamefont {Z.}~\bibnamefont {Zuo}}, \bibinfo {author}
  {\bibfnamefont {A.}~\bibnamefont {Bernard}}, \bibinfo {author} {\bibfnamefont
  {B.}~\bibnamefont {Hambrecht}}, \bibinfo {author} {\bibfnamefont
  {P.}~\bibnamefont {Lugan}}, \bibinfo {author} {\bibfnamefont
  {D.}~\bibnamefont {Clement}}, \bibinfo {author} {\bibfnamefont
  {L.}~\bibnamefont {Sanchez-Palencia}}, \bibinfo {author} {\bibfnamefont
  {P.}~\bibnamefont {Bouyer}},\ and\ \bibinfo {author} {\bibfnamefont
  {A.}~\bibnamefont {Aspect}},\ }\bibfield  {title} {\bibinfo {title} {Direct
  observation of {Anderson} localization of matter waves in a controlled
  disorder},\ }\href {https://doi.org/10.1038/nature07000} {\bibfield
  {journal} {\bibinfo  {journal} {Nature}\ }\textbf {\bibinfo {volume} {453}},\
  \bibinfo {pages} {891} (\bibinfo {year} {2008})}\BibitemShut {NoStop}%
\bibitem [{\citenamefont {Gadway}\ \emph {et~al.}(2011)\citenamefont {Gadway},
  \citenamefont {Pertot}, \citenamefont {Reeves}, \citenamefont {Vogt},\ and\
  \citenamefont {Schneble}}]{Gadway2011}%
  \BibitemOpen
  \bibfield  {author} {\bibinfo {author} {\bibfnamefont {B.}~\bibnamefont
  {Gadway}}, \bibinfo {author} {\bibfnamefont {D.}~\bibnamefont {Pertot}},
  \bibinfo {author} {\bibfnamefont {J.}~\bibnamefont {Reeves}}, \bibinfo
  {author} {\bibfnamefont {M.}~\bibnamefont {Vogt}},\ and\ \bibinfo {author}
  {\bibfnamefont {D.}~\bibnamefont {Schneble}},\ }\bibfield  {title} {\bibinfo
  {title} {Glassy behavior in a binary atomic mixture},\ }\href
  {https://doi.org/10.1103/PhysRevLett.107.145306} {\bibfield  {journal}
  {\bibinfo  {journal} {Physical Review Letters}\ }\textbf {\bibinfo {volume}
  {107}},\ \bibinfo {pages} {145306} (\bibinfo {year} {2011})}\BibitemShut
  {NoStop}%
\bibitem [{\citenamefont {Fallani}\ \emph {et~al.}(2007)\citenamefont
  {Fallani}, \citenamefont {Lye}, \citenamefont {Guarrera}, \citenamefont
  {Fort},\ and\ \citenamefont {Inguscio}}]{Fallani2007}%
  \BibitemOpen
  \bibfield  {author} {\bibinfo {author} {\bibfnamefont {L.}~\bibnamefont
  {Fallani}}, \bibinfo {author} {\bibfnamefont {J.~E.}\ \bibnamefont {Lye}},
  \bibinfo {author} {\bibfnamefont {V.}~\bibnamefont {Guarrera}}, \bibinfo
  {author} {\bibfnamefont {C.}~\bibnamefont {Fort}},\ and\ \bibinfo {author}
  {\bibfnamefont {M.}~\bibnamefont {Inguscio}},\ }\bibfield  {title} {\bibinfo
  {title} {Ultracold atoms in a disordered crystal of light: Towards a {Bose}
  glass},\ }\href {https://doi.org/10.1103/PhysRevLett.98.130404} {\bibfield
  {journal} {\bibinfo  {journal} {Physical Review Letters}\ }\textbf {\bibinfo
  {volume} {98}},\ \bibinfo {pages} {130404} (\bibinfo {year}
  {2007})}\BibitemShut {NoStop}%
\bibitem [{\citenamefont {Roati}\ \emph {et~al.}(2008)\citenamefont {Roati},
  \citenamefont {D/'Errico}, \citenamefont {Fallani}, \citenamefont {Fattori},
  \citenamefont {Fort}, \citenamefont {Zaccanti}, \citenamefont {Modugno},
  \citenamefont {Modugno},\ and\ \citenamefont {Inguscio}}]{Roati2008}%
  \BibitemOpen
  \bibfield  {author} {\bibinfo {author} {\bibfnamefont {G.}~\bibnamefont
  {Roati}}, \bibinfo {author} {\bibfnamefont {C.}~\bibnamefont {D/'Errico}},
  \bibinfo {author} {\bibfnamefont {L.}~\bibnamefont {Fallani}}, \bibinfo
  {author} {\bibfnamefont {M.}~\bibnamefont {Fattori}}, \bibinfo {author}
  {\bibfnamefont {C.}~\bibnamefont {Fort}}, \bibinfo {author} {\bibfnamefont
  {M.}~\bibnamefont {Zaccanti}}, \bibinfo {author} {\bibfnamefont
  {G.}~\bibnamefont {Modugno}}, \bibinfo {author} {\bibfnamefont
  {M.}~\bibnamefont {Modugno}},\ and\ \bibinfo {author} {\bibfnamefont
  {M.}~\bibnamefont {Inguscio}},\ }\bibfield  {title} {\bibinfo {title}
  {Anderson localization of a non-interacting {Bose--Einstein} condensate},\
  }\href {https://doi.org/10.1038/nature07071} {\bibfield  {journal} {\bibinfo
  {journal} {Nature}\ }\textbf {\bibinfo {volume} {453}},\ \bibinfo {pages}
  {895} (\bibinfo {year} {2008})}\BibitemShut {NoStop}%
\bibitem [{\citenamefont {Mitra}\ \emph {et~al.}(2018)\citenamefont {Mitra},
  \citenamefont {Brown}, \citenamefont {Guardado-Sanchez}, \citenamefont
  {Kondov}, \citenamefont {Devakul}, \citenamefont {Huse}, \citenamefont
  {Schauß},\ and\ \citenamefont {Bakr}}]{Mitra2018}%
  \BibitemOpen
  \bibfield  {author} {\bibinfo {author} {\bibfnamefont {D.}~\bibnamefont
  {Mitra}}, \bibinfo {author} {\bibfnamefont {P.~T.}\ \bibnamefont {Brown}},
  \bibinfo {author} {\bibfnamefont {E.}~\bibnamefont {Guardado-Sanchez}},
  \bibinfo {author} {\bibfnamefont {S.~S.}\ \bibnamefont {Kondov}}, \bibinfo
  {author} {\bibfnamefont {T.}~\bibnamefont {Devakul}}, \bibinfo {author}
  {\bibfnamefont {D.~A.}\ \bibnamefont {Huse}}, \bibinfo {author}
  {\bibfnamefont {P.}~\bibnamefont {Schauß}},\ and\ \bibinfo {author}
  {\bibfnamefont {W.~S.}\ \bibnamefont {Bakr}},\ }\bibfield  {title} {\bibinfo
  {title} {Quantum gas microscopy of an attractive {Fermi}-{Hubbard} system},\
  }\href {https://doi.org/10.1038/nphys4297} {\bibfield  {journal} {\bibinfo
  {journal} {Nature Physics}\ }\textbf {\bibinfo {volume} {14}},\ \bibinfo
  {pages} {173} (\bibinfo {year} {2018})}\BibitemShut {NoStop}%
\bibitem [{\citenamefont {Gall}\ \emph {et~al.}(2020)\citenamefont {Gall},
  \citenamefont {Chan}, \citenamefont {Wurz},\ and\ \citenamefont
  {K\"ohl}}]{Gall2020}%
  \BibitemOpen
  \bibfield  {author} {\bibinfo {author} {\bibfnamefont {M.}~\bibnamefont
  {Gall}}, \bibinfo {author} {\bibfnamefont {C.~F.}\ \bibnamefont {Chan}},
  \bibinfo {author} {\bibfnamefont {N.}~\bibnamefont {Wurz}},\ and\ \bibinfo
  {author} {\bibfnamefont {M.}~\bibnamefont {K\"ohl}},\ }\bibfield  {title}
  {\bibinfo {title} {Simulating a {Mott} insulator using attractive
  interaction},\ }\href {https://doi.org/10.1103/PhysRevLett.124.010403}
  {\bibfield  {journal} {\bibinfo  {journal} {Phys. Rev. Lett.}\ }\textbf
  {\bibinfo {volume} {124}},\ \bibinfo {pages} {010403} (\bibinfo {year}
  {2020})}\BibitemShut {NoStop}%
\bibitem [{\citenamefont {Prasad}\ \emph {et~al.}(2014)\citenamefont {Prasad},
  \citenamefont {Medhi},\ and\ \citenamefont {Shenoy}}]{Prasad2014}%
  \BibitemOpen
  \bibfield  {author} {\bibinfo {author} {\bibfnamefont {Y.}~\bibnamefont
  {Prasad}}, \bibinfo {author} {\bibfnamefont {A.}~\bibnamefont {Medhi}},\ and\
  \bibinfo {author} {\bibfnamefont {V.~B.}\ \bibnamefont {Shenoy}},\ }\bibfield
   {title} {\bibinfo {title} {Fermionic superfluid from a bilayer band
  insulator in an optical lattice},\ }\href
  {https://doi.org/10.1103/PhysRevA.89.043605} {\bibfield  {journal} {\bibinfo
  {journal} {Phys. Rev. A}\ }\textbf {\bibinfo {volume} {89}},\ \bibinfo
  {pages} {043605} (\bibinfo {year} {2014})}\BibitemShut {NoStop}%
\bibitem [{\citenamefont {Prasad}(2022)}]{Prasad2022}%
  \BibitemOpen
  \bibfield  {author} {\bibinfo {author} {\bibfnamefont {Y.}~\bibnamefont
  {Prasad}},\ }\bibfield  {title} {\bibinfo {title} {Finite-temperature study
  of correlations in a bilayer band insulator},\ }\href
  {https://doi.org/10.1103/PhysRevB.106.184506} {\bibfield  {journal} {\bibinfo
   {journal} {Phys. Rev. B}\ }\textbf {\bibinfo {volume} {106}},\ \bibinfo
  {pages} {184506} (\bibinfo {year} {2022})}\BibitemShut {NoStop}%
\bibitem [{\citenamefont {Lee}\ \emph {et~al.}(2014)\citenamefont {Lee},
  \citenamefont {Zhang}, \citenamefont {Jeschke},\ and\ \citenamefont
  {Valent\'{\i}}}]{Lee2014}%
  \BibitemOpen
  \bibfield  {author} {\bibinfo {author} {\bibfnamefont {H.}~\bibnamefont
  {Lee}}, \bibinfo {author} {\bibfnamefont {Y.-Z.}\ \bibnamefont {Zhang}},
  \bibinfo {author} {\bibfnamefont {H.~O.}\ \bibnamefont {Jeschke}},\ and\
  \bibinfo {author} {\bibfnamefont {R.}~\bibnamefont {Valent\'{\i}}},\
  }\bibfield  {title} {\bibinfo {title} {Competition between band and {Mott}
  insulators in the bilayer {Hubbard} model: A dynamical cluster approximation
  study},\ }\href {https://doi.org/10.1103/PhysRevB.89.035139} {\bibfield
  {journal} {\bibinfo  {journal} {Phys. Rev. B}\ }\textbf {\bibinfo {volume}
  {89}},\ \bibinfo {pages} {035139} (\bibinfo {year} {2014})}\BibitemShut
  {NoStop}%
\bibitem [{\citenamefont {Rüger}\ \emph {et~al.}(2014)\citenamefont {Rüger},
  \citenamefont {Tocchio}, \citenamefont {Valentí},\ and\ \citenamefont
  {Gros}}]{Ruger2014}%
  \BibitemOpen
  \bibfield  {author} {\bibinfo {author} {\bibfnamefont {R.}~\bibnamefont
  {Rüger}}, \bibinfo {author} {\bibfnamefont {L.~F.}\ \bibnamefont {Tocchio}},
  \bibinfo {author} {\bibfnamefont {R.}~\bibnamefont {Valentí}},\ and\
  \bibinfo {author} {\bibfnamefont {C.}~\bibnamefont {Gros}},\ }\bibfield
  {title} {\bibinfo {title} {The phase diagram of the square lattice bilayer
  {Hubbard} model: {A} variational {Monte Carlo} study},\ }\href
  {https://doi.org/10.1088/1367-2630/16/3/033010} {\bibfield  {journal}
  {\bibinfo  {journal} {New Journal of Physics}\ }\textbf {\bibinfo {volume}
  {16}},\ \bibinfo {pages} {033010} (\bibinfo {year} {2014})}\BibitemShut
  {NoStop}%
\bibitem [{\citenamefont {Huscroft}\ and\ \citenamefont
  {Scalettar}(1997)}]{Huscroft1997}%
  \BibitemOpen
  \bibfield  {author} {\bibinfo {author} {\bibfnamefont {C.}~\bibnamefont
  {Huscroft}}\ and\ \bibinfo {author} {\bibfnamefont {R.~T.}\ \bibnamefont
  {Scalettar}},\ }\bibfield  {title} {\bibinfo {title} {Effect of disorder on
  charge-density wave and superconducting order in the half-filled attractive
  {Hubbard} model},\ }\href {https://doi.org/10.1103/PhysRevB.55.1185}
  {\bibfield  {journal} {\bibinfo  {journal} {Phys. Rev. B}\ }\textbf {\bibinfo
  {volume} {55}},\ \bibinfo {pages} {1185} (\bibinfo {year}
  {1997})}\BibitemShut {NoStop}%
\bibitem [{\citenamefont {Blankenbecler}\ \emph {et~al.}(1981)\citenamefont
  {Blankenbecler}, \citenamefont {Scalapino},\ and\ \citenamefont
  {Sugar}}]{Blankenbecler1981}%
  \BibitemOpen
  \bibfield  {author} {\bibinfo {author} {\bibfnamefont {R.}~\bibnamefont
  {Blankenbecler}}, \bibinfo {author} {\bibfnamefont {D.~J.}\ \bibnamefont
  {Scalapino}},\ and\ \bibinfo {author} {\bibfnamefont {R.~L.}\ \bibnamefont
  {Sugar}},\ }\bibfield  {title} {\bibinfo {title} {{Monte Carlo} calculations
  of coupled boson--fermion systems. {I}},\ }\href
  {https://doi.org/10.1103/PhysRevD.24.2278} {\bibfield  {journal} {\bibinfo
  {journal} {Physical Review D}\ }\textbf {\bibinfo {volume} {24}},\ \bibinfo
  {pages} {2278} (\bibinfo {year} {1981})}\BibitemShut {NoStop}%
\bibitem [{\citenamefont {Santos}(2003)}]{Santos2003}%
  \BibitemOpen
  \bibfield  {author} {\bibinfo {author} {\bibfnamefont {R.~R.~d.}\
  \bibnamefont {Santos}},\ }\bibfield  {title} {\bibinfo {title} {Introduction
  to quantum {Monte Carlo} simulations for fermionic systems},\ }\href@noop {}
  {\bibfield  {journal} {\bibinfo  {journal} {Brazilian Journal of Physics}\
  }\textbf {\bibinfo {volume} {33}},\ \bibinfo {pages} {36} (\bibinfo {year}
  {2003})}\BibitemShut {NoStop}%
\bibitem [{\citenamefont {Huse}(1988)}]{Huse1988}%
  \BibitemOpen
  \bibfield  {author} {\bibinfo {author} {\bibfnamefont {D.~A.}\ \bibnamefont
  {Huse}},\ }\bibfield  {title} {\bibinfo {title} {Ground-state staggered
  magnetization of two--dimensional quantum {Heisenberg} antiferromagnets},\
  }\href {https://doi.org/10.1103/PhysRevB.37.2380} {\bibfield  {journal}
  {\bibinfo  {journal} {Physical Review B}\ }\textbf {\bibinfo {volume} {37}},\
  \bibinfo {pages} {2380} (\bibinfo {year} {1988})}\BibitemShut {NoStop}%
\bibitem [{\citenamefont {Fisher}\ \emph {et~al.}(1989)\citenamefont {Fisher},
  \citenamefont {Weichman}, \citenamefont {Grinstein},\ and\ \citenamefont
  {Fisher}}]{Fisher1989}%
  \BibitemOpen
  \bibfield  {author} {\bibinfo {author} {\bibfnamefont {M.~P.~A.}\
  \bibnamefont {Fisher}}, \bibinfo {author} {\bibfnamefont {P.~B.}\
  \bibnamefont {Weichman}}, \bibinfo {author} {\bibfnamefont {G.}~\bibnamefont
  {Grinstein}},\ and\ \bibinfo {author} {\bibfnamefont {D.~S.}\ \bibnamefont
  {Fisher}},\ }\bibfield  {title} {\bibinfo {title} {Boson localization and the
  superfluid--insulator transition},\ }\href
  {https://doi.org/10.1103/PhysRevB.40.546} {\bibfield  {journal} {\bibinfo
  {journal} {Physical Review B}\ }\textbf {\bibinfo {volume} {40}},\ \bibinfo
  {pages} {546} (\bibinfo {year} {1989})}\BibitemShut {NoStop}%
\bibitem [{\citenamefont {Mondaini}\ \emph {et~al.}(2015)\citenamefont
  {Mondaini}, \citenamefont {Nikoli\ifmmode~\acute{c}\else \'{c}\fi{}},\ and\
  \citenamefont {Rigol}}]{Mondaini2015}%
  \BibitemOpen
  \bibfield  {author} {\bibinfo {author} {\bibfnamefont {R.}~\bibnamefont
  {Mondaini}}, \bibinfo {author} {\bibfnamefont {P.}~\bibnamefont
  {Nikoli\ifmmode~\acute{c}\else \'{c}\fi{}}},\ and\ \bibinfo {author}
  {\bibfnamefont {M.}~\bibnamefont {Rigol}},\ }\bibfield  {title} {\bibinfo
  {title} {{Mott}--insulator to superconductor transition in a two--dimensional
  superlattice},\ }\href {https://doi.org/10.1103/PhysRevA.92.013601}
  {\bibfield  {journal} {\bibinfo  {journal} {Physical Review A}\ }\textbf
  {\bibinfo {volume} {92}},\ \bibinfo {pages} {013601} (\bibinfo {year}
  {2015})}\BibitemShut {NoStop}%
\bibitem [{\citenamefont {Bai}\ \emph {et~al.}(2009)\citenamefont {Bai},
  \citenamefont {Chen}, \citenamefont {Scalettar},\ and\ \citenamefont
  {Yamazaki}}]{Bai2009}%
  \BibitemOpen
  \bibfield  {author} {\bibinfo {author} {\bibfnamefont {Z.}~\bibnamefont
  {Bai}}, \bibinfo {author} {\bibfnamefont {W.}~\bibnamefont {Chen}}, \bibinfo
  {author} {\bibfnamefont {R.}~\bibnamefont {Scalettar}},\ and\ \bibinfo
  {author} {\bibfnamefont {I.}~\bibnamefont {Yamazaki}},\ }\bibfield  {title}
  {\bibinfo {title} {Numerical methods for quantum {Monte Carlo} simulations of
  the {Hubbard} model},\ }in\ \href@noop {} {\emph {\bibinfo {booktitle}
  {Multi-Scale Phenomena In Complex Fluids: Modeling, Analysis and Numerical
  Simulation}}}\ (\bibinfo  {publisher} {World Scientific},\ \bibinfo {year}
  {2009})\ pp.\ \bibinfo {pages} {1--110}\BibitemShut {NoStop}%
\end{thebibliography}%

\end{document}